\documentclass[]{JHEP3}

\usepackage{multicol}
\usepackage{epsf,amsmath,amssymb}

\newcommand{\nb}{\nonumber}

\title{Pseudoscalar Higgs boson production associated with a single bottom quark at hadron colliders}

\author{{ Hou Hong-Sheng$^{b}$, Ma Wen-Gan$^{a,b}$, Zhang Ren-You$^{b}$, Sun Yan-Bin$^{b}$ and Wu Peng$^{b}$}\\
{\small $^{a}$CCAST (World Laboratory), P.O.Box 8730, Beijing,
100080, People's Republic of China} \\
{\small $^{b}$Department of Modern Physics, University of Science
and Technology of China (USTC),}
{\small       Hefei, Anhui 230027, People's Republic of China}\\
{\it Email: hhsheng@mail.ustc.edu.cn, mawg@ustc.edu.cn,
zhangry@mail.ustc.edu.cn, sunyb@mail.ustc.edu.cn, and
lily@mail.ustc.edu.cn}}

\abstract{ We compute the complete next-to-leading order (NLO)
SUSY-QCD corrections for the associated production of a
pseudoscalar Higgs boson with a bottom quark via bottom-gluon
fusion at the CERN Large Hadron Collider (LHC) and the Fermilab
Tevatron. We find that the NLO QCD correction in the MSSM reaches
$40\%\sim50\%$ at the LHC and $45\%\sim80\%$ at the Tevatron in
our chosen parameter space.}


\begin{document}

\section{Introduction}
The standard model (SM) \cite{sm}, with the Higgs mechanism
manifested by a single complex scalar weak isospin doublet that
spontaneously break the electrowak gauge symmetry, is believed to
be an incomplete description of nature. There exists the problem
of the quadratically divergent contributions to the corrections to
the Higgs boson mass. This is the so-called naturalness problem of
the SM. One of the good methods to solve this problem is to make
supersymmetric (SUSY) extensions to the SM. The quadratic
divergences of the Higgs mass can be cancelled by loop diagrams
involving the supersymmetric partners of the SM particles exactly.
The most attractive supersymmetric extension of the SM is the
minimal supersymmetric standard model (MSSM) \cite{mssm-1,mssm-2}.
Any enlargement of the sector beyond the single $SU(2)_{L}$ Higgs
doublet of SM, with two or more doublets as required in
supersymmetric theory, necessarily involves new physical
particles. In the MSSM, there are two Higgs doublets $H_1^0$ and
$H_2^0$ to give masses to down- and up-type fermions. The Higgs
sector consists of three neutral Higgs bosons, one $CP$-odd
particle ($A^0$), two $CP$-even particles ($h^0$ and $H^0$), and a
pair of charged Higgs bosons ($H^{\pm}$).

The discovery of the additional heavy Higgs bosons will help us to
probe the contents of the supersymmetric Higgs sector. Until now
all supersymmetric Higgs bosons haven't been directly explored
yet, only LEP2 group presents the strongest lower mass limits of
$91.0~GeV$ and $91.9~GeV$ for the light CP-even and the CP-odd
neutral Higgs bosons $h^{0}$ and $A^{0}$\cite{Lep2}, respectively.
The CERN Large Hadron Collider (LHC), which is a proton-proton
collider with $\sqrt{S}=14~TeV$ and a luminosity of
$100~\rm{fb}^{-1}$ per year, has been designed specifically to
continue finding Higgs bosons. In the MSSM theory with a large
value of $\tan \beta$($\tan{\beta}=v_2/v_1$, $v_1$ and $v_2$ are
the vacuum expectation values of the two Higgs boson fields
$H_1^0$ and $H_2^0$, respectively), the strength of the
$A^0b\bar{b}$ coupling increases greatly. The pseudoscalar Higgs
boson $A^0$ could be produced with a substantial rate at the LHC
either in association with bottom quarks [$q\bar{q},gg \to
b\bar{b}A^0$ or $gb(\bar{b}) \to b (\bar{b})A^0$] or through $gg
\to A^0$ when $b$ or $\tilde{b}$ loops dominate, provided that
$\tan \beta$ is large enough\cite{conway}. Ref.\cite{kauffman1}
presents the calculation of the total cross section of the $A^0$
plus two jets at the lowest order. The total cross section for the
inclusive production of $A^0$ has been calculated at NLO in
Ref.\cite{kauffman2} and at NNLO in
Refs.\cite{anastasiou}\cite{harlander}, where the authors use the
effective lagrangian for the interaction of $A^0$ Higgs boson with
the gluons and neglect the contribution of bottom quark loop so
that their result is only for small and moderate value of $\tan
\beta$.

Because the high-$p_T$ bottom quark can be tagged with reasonably
high efficiency, the observation of a bottom quark with high $p_T$
can reduce the backgrounds of the $A^0$ Higgs boson production.
The $A^0$ Higgs boson production associated with bottom quark can
occur via tree-level subprocess $gb(\bar{b}) \to
b(\bar{b})A^0$\cite{zhu}, where the initial bottom quark resides
in the proton sea. The cross section for the production of Higgs
boson $h^0$ and a single high $p_T$ bottom quark via subprocess
$gb(\bar{b}) \to b(\bar{b}) h^0$ has been studied in both QCD and
SUSY-QCD at NLO \cite{campbell1,hou,yang}. However, it is pointed
out in Ref.\cite{rainwater} that the calculation of $b\bar{b} \to
h^0$ may overestimate the inclusive cross section with the
introduction of conventional $b$ densities, due to crude
approximations inherent in the kinematics, which give rise to
large bottom quark mass and phase space effect. But in
Ref.\cite{maltoni}, it is shown that the bottom parton approach is
valid by choosing the appropriate factorization scale for the
process $b\bar{b} \to h^0$ with $\mu_f=m_h/4$ rather than $m_h$.
In Ref.\cite{plehn}, the author computed the NLO contributions to
the inclusive cross section $pp \to tH^-$ via the subprocess $bg
\to tH^-$ and shown the bottom parton approach is valid for this
process by choosing the factorization scale with about $\mu_f \sim
m_{\rm av}/3=(m_t+m_{H^-})/6$. In our calculation, we choose the
factorization scale as $\mu_f=m_{A}/4$ when we use the bottom
parton approximation.

In this paper, we calculated the cross section for the associated
production of the $A^0$ Higgs boson and a single high-$p_T$ bottom
quark via $bg \to A^0b(\bar{b}g \to A^0\bar{b})$ in the MSSM at
the Tevatron and LHC including the NLO QCD corrections. The
structure of this paper is as follow: In Sec. 2, we discuss the LO
results of the subprocess $bg \to A^0b$. In Sec. 3, we present the
calculations of the NLO QCD corrections. In Sec. 4, the numerical
results and conclusions are presented.

\section{The leading order cross section}
Since the cross sections for the subprocess $bg \to A^0b$ and its
charge-conjugate subprocess $\bar{b}g \to A^0\bar{b}$ in the
CP-conserved MSSM are same, we present only the calculation of the
subprocess $b(p_1)g(p_2) \to A^0(k_3)b(k_4)$ here (where $p_{1,2}$
and $k_{3,4}$ represent the four-momentum of the incoming partons
and the outgoing particles, respectively.). The subprocess $bg \to
A^0b$ can occur through both s-channel and t-channel as shown in
Fig.1(A-B). So we divide the tree-level amplitude into two parts
and denote it as
\begin{eqnarray}
M^0=M_0^{(s)}+M_0^{(t)},
\end{eqnarray}
where $M_0^{(s)}$ and $M_0^{(t)}$ represent the amplitudes arising
from the s-channel diagram shown in Fig.1(A) and the t-channel
diagram shown in Fig.1(B) at the tree-level, respectively. \\

\FIGURE{
  \leavevmode
  \begin{tabular}{cc}
    \epsfxsize=30em
      \epsffile[110 600 440 670]{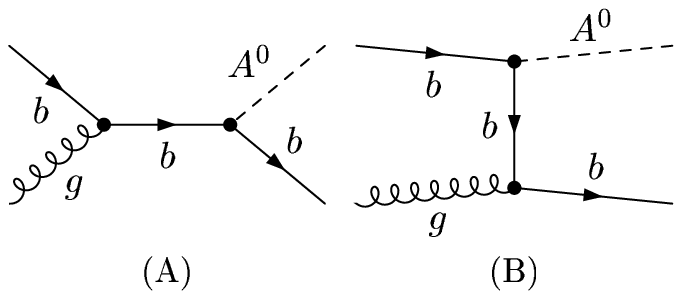}
  \end{tabular}
  \caption[]{\label{fig::fig1}\sloppy
    Leading order Feynman diagrams for the subprocess of
$bg \to A^0b$ }
  }

The explicit expressions for the amplitudes $M_0^{(s)}$ and
$M_0^{(t)}$ can be written as
\begin{eqnarray}
M_0^{(s)}&=&  \frac{g_s(\mu_r)V_{Abb}(\mu_r)}{\hat{s}}
   \bar{u}_i(k_4) \gamma_5( \rlap/p_1 + \rlap/p_2) \gamma_{\nu} u_j(p_1)\epsilon^a_{\nu}(p_2)T^a_{ij},  \nb \\
M_0^{(t)}&=& \frac{g_s(\mu_r) V_{Abb }(\mu_r)}{\hat{t}}
   \bar{u}_i(k_4) \gamma_{\nu} (\rlap/p_1 - \rlap/k_3)
   \gamma_5 u_j(p_1)\epsilon^a_{\nu}(p_2)T^a_{ij},
\end{eqnarray}
where $\hat{s}=(p_1+p_2)^2$, $\hat{t}=(p_1-k_3)^2$ and
$\hat{u}=(p_1-k_4)^2$ are the usual Mandelstam variables. $\mu_r$
is the renormalization scale, $g_s(\mu_r)$ is the running strong
coupling strength and $T^a$ is the $SU(3)$ color matrix.
$V_{Abb}(\mu_r)\gamma_5$ is the Yukawa coupling between $A^0$
Higgs boson and bottom quarks. In MSSM, $V_{Abb}(\mu_r)$ is given
as
\begin{eqnarray}
\label{vertex1}
 V_{Abb}(\mu_r)=- \frac{g_w \overline{m}_b(\mu_r)
\tan\beta}{2 m_W}
\end{eqnarray}
$\overline{m}_b(\mu_r)$ is the $\overline{\rm MS}$ running mass of
the bottom quark. It is well known that we should use the running
mass rather than the pole mass when evaluation of the Yukawa
coupling, because the pole mass Yukawa coupling will yield a huge
overestimate of the cross section. Throughout our evaluation we
neglect the bottom quark mass except in the Yukawa couplings. This
corresponds to the simplified Aivazis-Collins-Olness-Tung(ACOT)
scheme\cite{aivazis}. In any diagram in which the bottom quark is
an initial-state parton, the bottom quark mass may be neglected
without any loss of accuracy.

Then the lowest order cross section for the subprocess $bg \to
A^0b$ in the MSSM is obtained by using the following formula:
\begin{eqnarray}
\label{folding} \hat{\sigma}^0(\hat{s}, bg \to A^0b) = \frac{1}{16
\pi \hat{s}^2} \int_{\hat{t}_{min}}^{\hat{t}_{max}} d\hat{t}~
\overline{\sum} |M^0|^2,
\end{eqnarray}
where $\hat{t}_{max}=0$ and $\hat{t}_{min}=m_A^2-\hat{s}$. The
summation is taken over the spins and colors of initial and final
states, and the bar over the summation recalls averaging over the
spins and colors of initial partons.

\section{NLO QCD corrections}
The NLO QCD contributions to the subprocess $bg \to A^0b$ can be
separated into two parts: the virtual corrections arising from
loop diagrams and the real gluon emission corrections.

\subsection{Virtual Corrections}
The virtual corrections in the MSSM to $bg \to A^0b$ consist of
self-energy, vertex and box diagrams which are shown in Figs.2-3.
Fig.2 shows the one-loop diagrams of the SM-like QCD corrections
from quarks and gluons, and Fig.3 presents the one-loop diagrams
of the SUSY QCD corrections from squarks and gluinos. There exist
both ultraviolet(UV) and soft/collinear infrared(IR) singularities
in the amplitude from the SM-like diagrams in Fig.2, and the
amplitude part from SUSY QCD diagrams(Fig.3) only contains UV
singularities. In our calculation, we adopt the 't Hooft-Feynman
gauge and all the divergences are regularized by using dimensional
regularization method in $d=4-2 \epsilon$ dimensions.

\FIGURE{
  \leavevmode
  \begin{tabular}{cc}
    \epsfxsize=25em
      \epsffile[70 180 430 700]{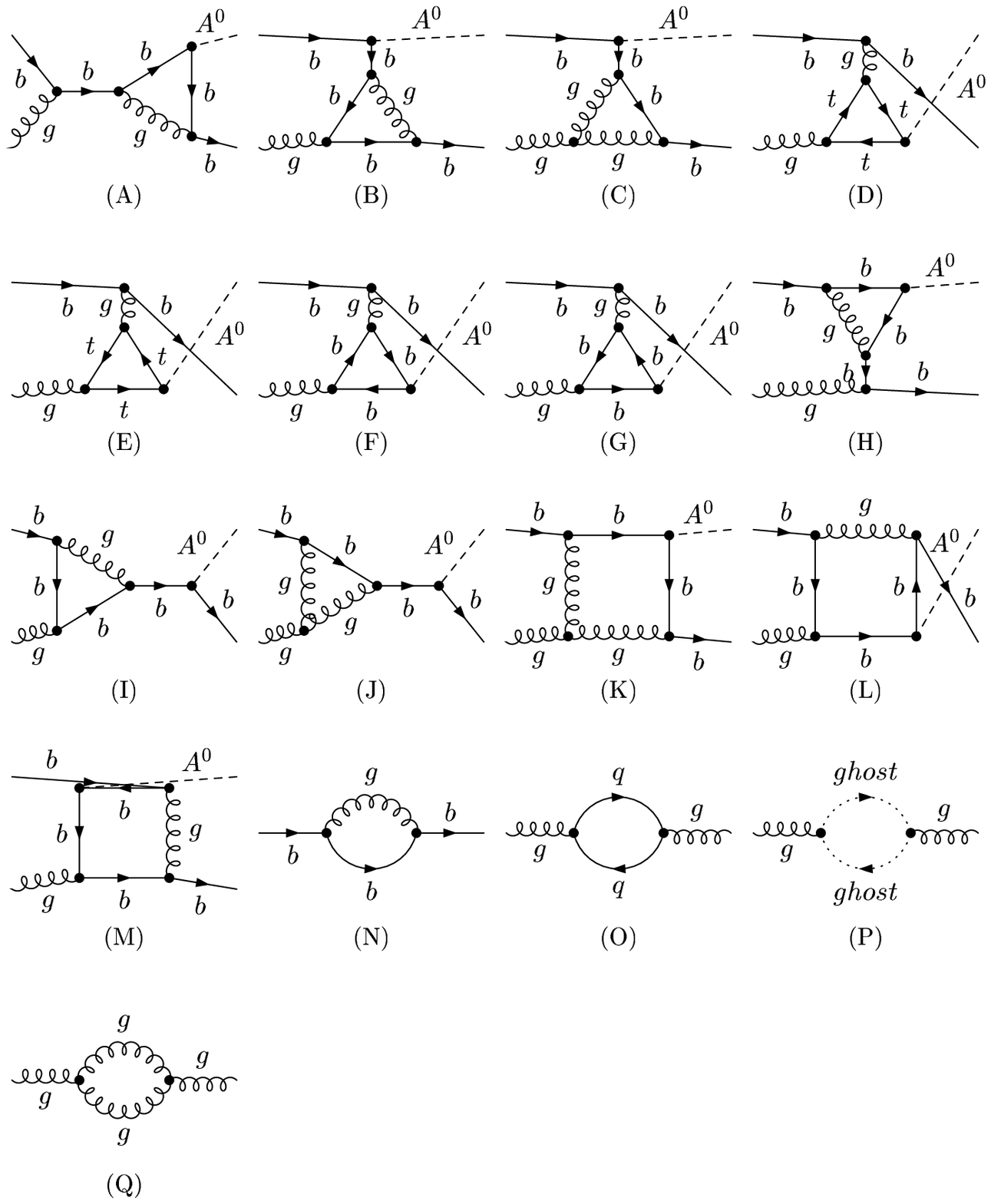}
  \end{tabular}
  \caption[]{\label{fig::fig2}\sloppy
    Virtual one-loop Feynman diagrams for the subprocess
of $bg \to A^0b$ of the SM-like QCD corrections. }
  }

\FIGURE{
  \leavevmode
  \begin{tabular}{cc}
    \epsfxsize=25em
      \epsffile[70 180 430 700]{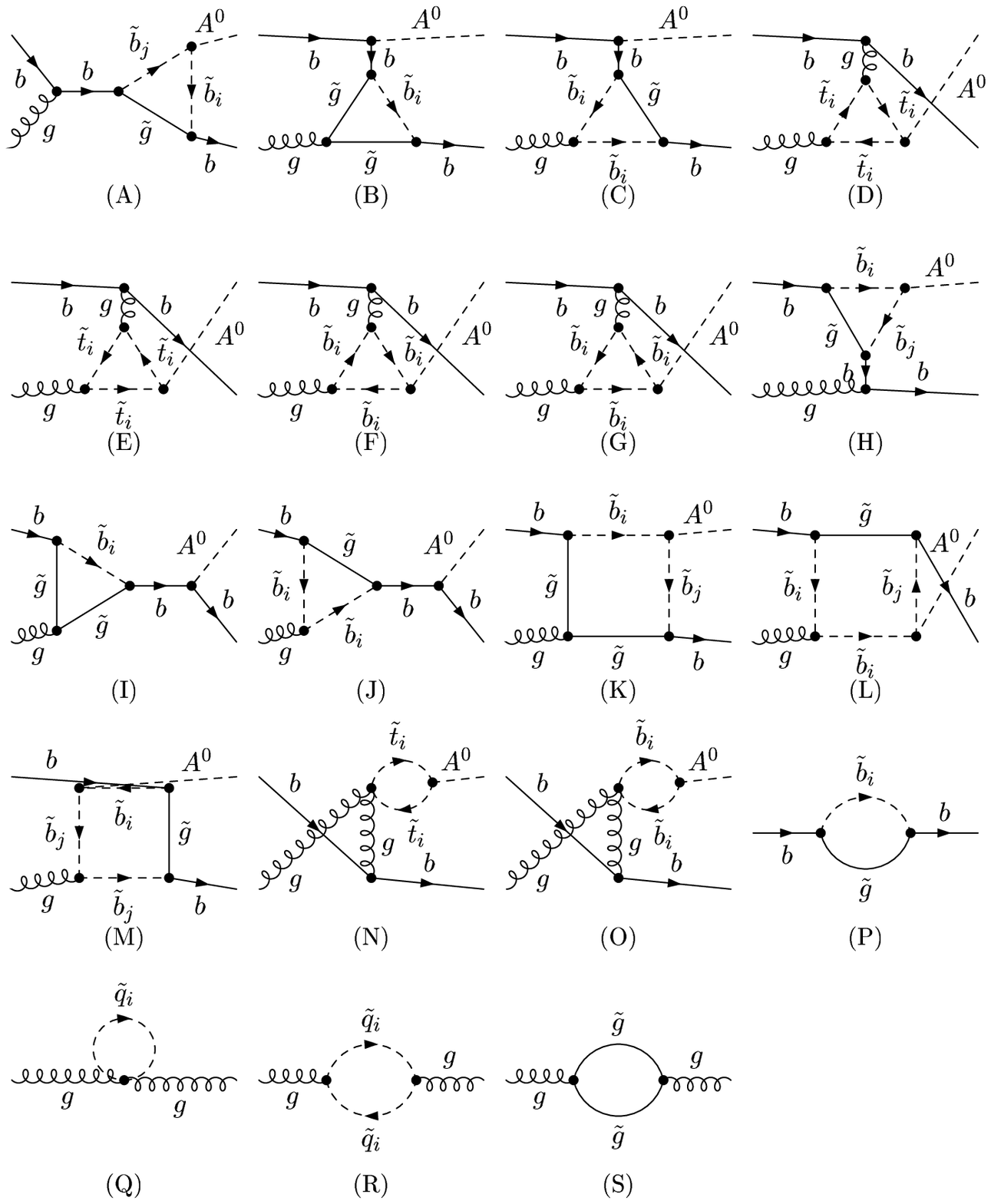}
  \end{tabular}
  \caption[]{\label{fig::fig3}\sloppy
    Virtual one-loop Feynman diagrams for the subprocess
of $bg \to A^0b$ of the SUSY QCD corrections. }
  }

In order to remove the UV divergences, we need to renormalize the
wave functions of the external fields, the strong coupling and the
$A^0-b-\bar{b}$ Yukawa coupling. For the renormalization of the
strong and Yukawa couplings, we employ the modified Minimal
Subtraction ($\overline{\rm MS}$) scheme. The relevant
renormalization constants in this work are defined same as those
in Ref.\cite{hou}.
\par
The virtual corrections to the cross section can be written as
\begin{eqnarray}
\hat{\sigma}^{V}(\hat{s}, bg \to A^0b) = \frac{1}{16 \pi
\hat{s}^2} \int_{\hat{t}_{min}}^{\hat{t}_{max}} d\hat{t}~ 2 Re
\overline{\sum} [(M^{V})^{\dagger} M^0],
\end{eqnarray}
with $\hat{t}_{max}=0$ and $\hat{t}_{min}=m_A^2-\hat{s}$ and again
the summation with bar means the same operations as appeared in
Eq.(\ref{folding}). $M^{V}$ is the renormalized amplitude for
virtual corrections.

After the renormalization procedure, $\hat{\sigma}^{V}$ is
UV-finite. Nevertheless, it still contains the soft/collinear IR
singularities
\begin{eqnarray}
\label{virtual cross section}
d\hat{\sigma}^V|_{IR}=\left[\frac{\alpha_s}{2 \pi}
\frac{\Gamma(1-\epsilon)}{\Gamma(1-2 \epsilon)}\left(\frac{4 \pi
\mu_r^2}{\hat{s}}\right)^{\epsilon}\right]d\hat{\sigma}^0
\left(\frac{A^V_2}{\epsilon^2}+\frac{A^V_1}{\epsilon} \right),
\end{eqnarray}
where
\begin{eqnarray}
A^V_2&=&-\frac{17}{3}, \nb \\
A^V_1&=&-\frac{47}{6}+3 \ln \frac{-\hat{t}}{\hat{s}-m_{A}^2}
-\frac{1}{3} \ln \frac{-\hat{u}}{\hat{s}-m_{A}^2}.
\end{eqnarray}

 The soft divergences can be cancelled by
adding with the soft real gluon emission corrections, and the
remaining collinear divergences are absorbed into the parton
distribution functions, which will be discussed in the next
subsection.

\subsection{Real gluon emission corrections}
The $O(\alpha_s)$ corrections to $bg \to A^0b$ due to real gluon
emission (shown in Fig.4) give the origin of IR singularities
which cancel exactly the analogous singularities present in the
$O(\alpha_s)$ virtual corrections mentioned in above subsection.
These singularities can be either of soft or collinear nature and
can be conveniently isolated by slicing the $bg \to A^0b+g$ phase
space into different regions defined by suitable cutoffs, a method
which goes under the general name of Phase Space Slicing(PPS).

\FIGURE{
  \leavevmode
  \begin{tabular}{cc}
    \epsfxsize=25em
      \epsffile[70 490 430 700]{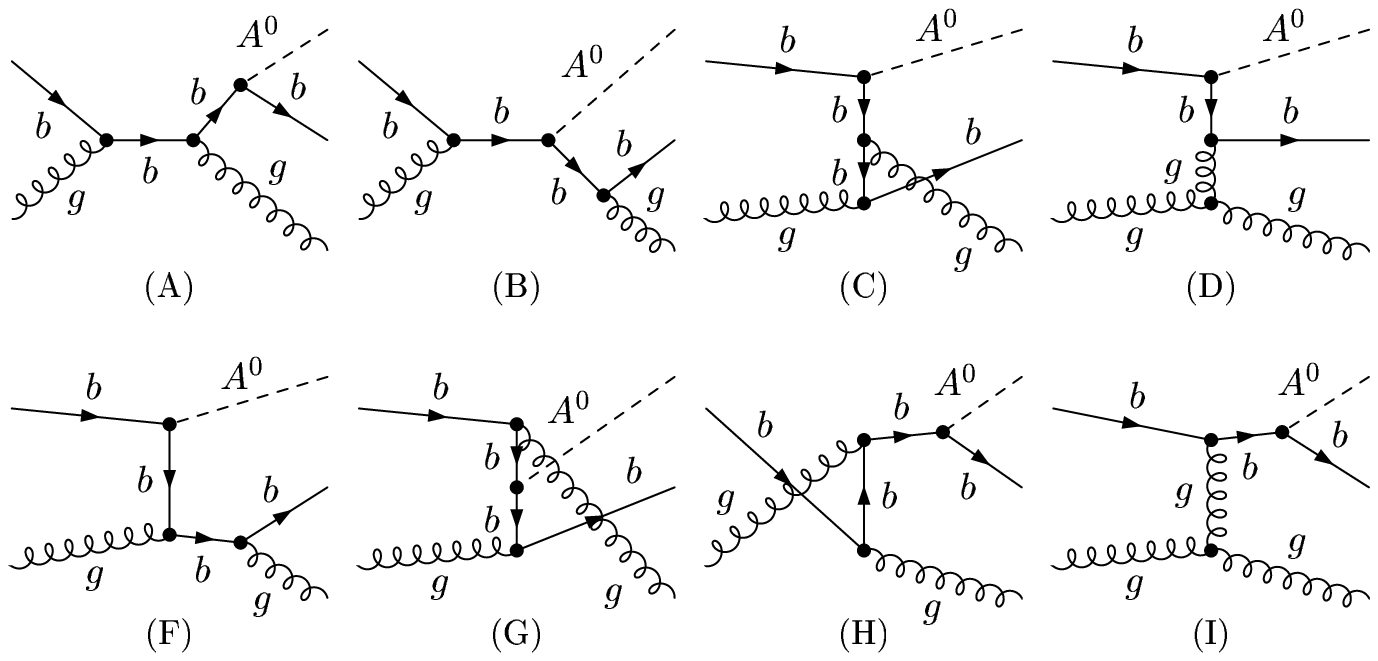}
  \end{tabular}
  \caption[]{\label{fig::fig4}\sloppy
    Feynman diagrams for the subprocess of $bg \to A^0bg$
with a real gluon emission. }
  }

In this paper, we calculate the cross section for the $2 \to 3$
process
\begin{eqnarray}
b(p_1)+g(p_2) \to A^0(k_3)+b(k_4)+g(k_5),
\end{eqnarray}
adopting the method named two cutoff phase space slicing
method\cite{Harris}. We define the invariants
\begin{eqnarray}
\hat{s}&=&(p_1+p_2)^2,~~\hat{t}=(p_1-k_3)^2,~~\hat{u}=(p_1-k_4)^2,
\nb \\
\hat{s}_{45}&=&(k_4+k_5)^2,~~\hat{t}_{15}=(p_1-k_5)^2,~~
\hat{t}_{25}=(p_2-k_5)^2,~~\hat{t}_{45}=(k_4-k_5)^2,
\end{eqnarray}
and describe this method briefly as follows. Firstly, by
introducing an arbitrary small soft cutoff $\delta_s$ we separate
the $2 \to 3$ phase space into two regions, according to whether
the energy of the emitted gluon is soft, i.e. $E_5 \leq
\delta_s\sqrt{\hat{s}}/2$, or hard, i.e. $E_5 >
\delta_s\sqrt{\hat{s}}/2$. The partonic real cross section can be
written as
\begin{eqnarray}
\hat{\sigma}^R(bg \to A^0bg)=\hat{\sigma}^S(bg \to
A^0bg)+\hat{\sigma}^H(bg \to A^0bg),
\end{eqnarray}
where $\hat{\sigma}^S$ is obtained by integrating over the soft
region of the emitted gluon phase space. $\hat{\sigma}^S$ contains
all the soft IR singularities. Secondly, to isolate the remaining
collinear singularities from $\hat{\sigma}^H$, we further
decompose $\hat{\sigma}^H$ into a sum of hard-collinear (HC) and
hard-non-collinear ($\overline{\rm HC}$) terms by introducing
another cutoff $\delta_c$ named collinear cutoff
\begin{eqnarray}
\hat{\sigma}^H(bg \to A^0bg)=\hat{\sigma}^{\rm HC}(bg \to
A^0bg)+\hat{\sigma}^{\overline{\rm HC}}(bg \to A^0bg).
\end{eqnarray}
The HC regions of the phase space are those where any invariant
$t_{15},t_{25},t_{45}$ becomes smaller in magnitude than $\delta_c
\hat{s} $, in collinear condition, while at the same time the
emitted gluon remains hard. $\hat{\sigma}^{\rm HC}$ contains the
collinear divergences. In the soft and HC region, $\hat{\sigma}^S$
and $\hat{\sigma}^{\rm HC}$ can be obtained by performing the
phase space integration in $d$-dimension analytically. In the
$\overline{\rm HC}$ region, $\hat{\sigma}^{\overline{\rm HC}}$ is
finite and may be evaluated in four dimensions using standard
Monte Carlo techniques\cite{Lepage}. The cross sections,
$\hat{\sigma}^S$, $\hat{\sigma}^{\rm HC}$ and
$\hat{\sigma}^{\overline{\rm HC}}$, depend on the two arbitrary
parameters, $\delta_s$ and $\delta_c$. However, in the total real
gluon emission hadronic cross section $\hat{\sigma}^R$, after mass
factorization, the dependence on these arbitrary cutoffs cancels,
as will be explicitly shown in Sec. 4. This constitutes an
important check of our calculation.
\par
The soft region of the $bg \to A^0b+g$ phase space is defined by
\begin{eqnarray}
0<E_5 \leq \delta_s\sqrt{\hat{s}}/2
\end{eqnarray}
\par
The differential cross section in the soft region is given as
\begin{eqnarray}
\label{soft cross section} d\hat{\sigma}^S=d\hat{\sigma}^0
\left[\frac{\alpha_s}{2 \pi} \frac{\Gamma(1-\epsilon)}{\Gamma(1-2
\epsilon)}\left(\frac{4 \pi
\mu_r^2}{\hat{s}}\right)^{\epsilon}\right]
\left(\frac{A^S_2}{\epsilon^2}+\frac{A^S_1}{\epsilon}+A^S_0
\right),
\end{eqnarray}
with
\begin{eqnarray}
A^S_2&=&\frac{17}{3}, \nb \\
A^S_1&=&-\frac{34}{3} \ln \delta_s - 3 \ln
\frac{-\hat{t}}{\hat{s}-m_{A}^2}
+\frac{1}{3} \ln \frac{-\hat{u}}{\hat{s}-m_{A}^2}, \nb \\
A^S_0&=& \frac{34}{3} \ln^2 \delta_s+ 6 \ln \delta_s \ln
\frac{-\hat{t}}{\hat{s}-m_{A}^2}+ \frac{3}{2} \ln^2
\frac{-\hat{t}}{\hat{s}-m_{A}^2}  \nb \\
&-& \frac{2}{3} \ln \delta_s \ln \frac{-\hat{u}}{\hat{s}-m_{A}^2}-
\frac{1}{6} \ln^2 \frac{-\hat{u}}{\hat{s}-m_{A}^2}-\frac{1}{3}
Li_2[\frac{-\hat{t}}{\hat{s}-m_{A}^2}]+3
Li_2[\frac{-\hat{u}}{\hat{s}-m_{A}^2}].
\end{eqnarray}
\par
In the limit where two of the partons are collinear, the three
body phase space is greatly simplified. And in the same limit, the
leading pole approximation of the matrix element is valid.
According to whether the collinear singularities are initial or
final state in origin, we separate $\hat{\sigma}^{\rm HC}$ into
two pieces
\begin{eqnarray}
\hat{\sigma}^{\rm HC}=\hat{\sigma}_i^{\rm HC}+\hat{\sigma}_f^{\rm
HC}.
\end{eqnarray}
$\hat{\sigma}_i^{\rm HC}$ is the cross section arising from the
case that the emitted gluon is collinear to the initial partons,
$0 \leq t_{15},t_{25} \leq \delta_c \hat{s}$. And
$\hat{\sigma}_f^{\rm HC}$ arises from the case that the emitted
gluon is collinear to the final parton, $0 \leq t_{45} \leq
\delta_c \hat{s}$.
\par
The cross section $\hat{\sigma}_f^{\rm HC}$ can be written as
\begin{eqnarray}
\label{final collinear cross section} d\hat{\sigma}_f^{\rm
HC}=d\hat{\sigma}^0 \left[\frac{\alpha_s}{2 \pi}
\frac{\Gamma(1-\epsilon)}{\Gamma(1-2 \epsilon)}\left(\frac{4 \pi
\mu_r^2}{\hat{s}}\right)^{\epsilon}\right] \left(\frac{A^{b \to bg
}_1}{\epsilon}+A^{b \to bg}_0 \right),
\end{eqnarray}
where
\begin{eqnarray}
A^{b \to bg }_1&=& C_F(3/2+2\ln\delta_s), \nb \\
A^{b \to bg }_0&=& C_F[7/2-\pi^2/3-\ln^2\delta_s-\ln\delta_c(3/2+2
\ln\delta_s )].
\end{eqnarray}
\par
The cross section $\sigma_i^{\rm HC}$ can be written as
\begin{eqnarray}
\label{initial collinear} d\sigma_i^{\rm HC}&=&d\hat{\sigma}^0
\left[\frac{\alpha_s}{2 \pi} \frac{\Gamma(1-\epsilon)}{\Gamma(1-2
\epsilon)}\left(\frac{4 \pi
\mu_r^2}{\hat{s}}\right)^{\epsilon}\right]
(-\frac{1}{\epsilon})\delta_c^{-\epsilon}
[P_{bb}(z,\epsilon)G_{b/A}(x_1/z)G_{g/B}(x_2) \nb \\
&+&P_{gg}(z,\epsilon)G_{g/A}(x_1/z)G_{b/B}(x_2)
+(x_1\leftrightarrow
x_2)]\frac{dz}{z}(\frac{1-z}{z})^{-\epsilon}dx_1dx_2.
\end{eqnarray}
where $G_{b,g/A,B}(x)$ is the bare parton distribution function,
$P_{bb}(z,\epsilon)$ and $P_{gg}(z,\epsilon)$ are the
$d$-dimensional unregulated ($z<1$) splitting function related to
the usual Altarelli-Parisi splitting kernels\cite{Altarelli}.
$P_{ii}(z,\epsilon)~~(i=b,g)$ can be written explicitly as
\begin{eqnarray}
P_{ii}(z,\epsilon)&=&P_{ii}(z)+ \epsilon P'_{ii}(z)~~~~~~~~(i=b,g), \nb \\
P_{bb}(z)&=&C_F \frac{1+z^2}{1-z},~~~~~~~~ P'_{bb}(z)=-C_F (1-z)
\nb \\
P_{gg}(z)&=&2 N[\frac{z}{1-z}+\frac{1-z}{z}+z(1-z)],~~~~~~~~
P'_{gg}(z)=0.
\end{eqnarray}
with $N=3$ and $C_F=4/3$.
\par
 In order to factorize the collinear
singularity of $\sigma_i^{\rm HC}$ into the parton distribution
function, we introduce a scale dependent parton distribution
function using the $\overline{\rm MS}$ convention:
\begin{eqnarray}
G_{i/A}(x,\mu_f)=G_{i/A}(x)+(-\frac{1}{\epsilon})\left[\frac{\alpha_s}{2
\pi} \frac{\Gamma(1-\epsilon)}{\Gamma(1-2 \epsilon)}\left(\frac{4
\pi
\mu_r^2}{\mu_f^2}\right)^{\epsilon}\right]\int^1_z\frac{dz}{z}P_{ii}(z)G_{i/A}(x/z),\nb
\\ (i=b,g).
\end{eqnarray}
By using above definition, we replace $G_{g,b/A,B}$ in
Eq.(\ref{initial collinear}) and the expression for the initial
state collinear contribution at $O(\alpha_s)$ order is
\begin{eqnarray}
\label{initial collinear cross section} d\sigma_i^{\rm
HC}&=&d\hat{\sigma}^0 \left[\frac{\alpha_s}{2 \pi}
\frac{\Gamma(1-\epsilon)}{\Gamma(1-2 \epsilon)}\left(\frac{4 \pi
\mu_r^2}{\hat{s}}\right)^{\epsilon}\right]\{
\tilde{G}_{g/A}(x_1,\mu_f)G_{b/B}(x_2,\mu_f)+G_{g/A}(x_1,\mu_f)\tilde{G}_{b/B}(x_2,\mu_f)
\nb \\
&+& \sum_{\alpha=g,b}[\frac{A_1^{sc}(\alpha \to \alpha
g)}{\epsilon}+A_0^{sc}(\alpha \to \alpha
g)]G_{g/A}(x_1,\mu_f)G_{b/B}(x_2,\mu_f) \nb \\
&+& (x_1 \leftrightarrow x_2)\}dx_1dx_2,
\end{eqnarray}
where
\begin{eqnarray}
A_1^{sc}(b \to bg)&=&C_F(2 \ln \delta_s+3/2), \nb \\
A_1^{sc}(g \to gg)&=&2 N \ln \delta_s + (11 N -2 n_f)/6 , \nb \\
A_0^{sc}&=&A_1^{sc} \ln(\frac{\hat{s}}{\mu_f^2}).
\end{eqnarray}
And
\begin{eqnarray}
\tilde{G}_{\alpha/A,B}(x,\mu_f)=\int^{1-\delta_s}_x
\frac{dy}{y}G_{\alpha/A,B}(x/y,\mu_f)\tilde{P}_{\alpha \alpha}(y),
~~~~(\alpha=g,b),
\end{eqnarray}
with
\begin{eqnarray}
\tilde{P}_{\alpha \alpha}(y)=P_{\alpha \alpha}
\ln(\delta_c\frac{1-y}{y}\frac{\hat{s}}{\mu_f^2})-P'_{\alpha
\alpha}(y,),~~~~(\alpha=g,b).
\end{eqnarray}
We can observe that the sum of the soft (Eq.(\ref{soft cross
section})), collinear(Eq.(\ref{final collinear cross
section}),(\ref{initial collinear cross section})), and
ultraviolet renormalized virtual correction (Eq.(\ref{virtual
cross section})) terms is finite, i.e.,
\begin{eqnarray}
A^S_2&+&A^V_2=0, \nb \\
A^S_1&+&A^V_1+A_1^{b \to bg}+A_1^{sc}(b\to bg)+A_1^{sc}(g\to
gg)=0.
\end{eqnarray}
The final result for the $O(\alpha_s)$ correction consists of two
contributions to the cross section: a two-body term $\sigma^{(2)}$
and a three-body term $\sigma^{(3)}$.
\begin{eqnarray}
\sigma^{(2)}&=&\frac{\alpha_s}{2 \pi} \int dx_1dx_2d\hat{\sigma}^0
\{ G_{g/A}(x_1,\mu_f)G_{b/B}(x_2,\mu_f)[A^S_0+A^V_0+A_0^{b \to
bg}+A_0^{sc}(b\to bg)+A_0^{sc}(g\to gg)] \nb \\
&+&\tilde{G}_{g/A}(x_1,\mu_f)G_{b/B}(x_2,\mu_f)+G_{g/A}(x_1,\mu_f)\tilde{G}_{b/B}(x_2,\mu_f)+(x_1
\leftrightarrow x_2 ) \}.
\end{eqnarray}
And
\begin{eqnarray}
\sigma^{(3)}=\int dx_1dx_2
[G_{g/A}(x_1,\mu_f)G_{b/B}(x_2,\mu_f)+(x_1 \leftrightarrow x_2
)]d\hat{\sigma}^{(3)},
\end{eqnarray}
with the hard-non-collinear partonic cross section given by
\begin{eqnarray}
d\hat{\sigma}^{(3)}=\frac{1}{2\hat{s}_{12}} \int_{\overline{\rm
HC}}\overline{\sum}|M_3(bg \to A^0bg)|^2 d \Gamma_3.
\end{eqnarray}
Finally, the NLO total cross section for $pp$(or $p\bar{p}) \to
bA^0+X$ is
\begin{eqnarray}
\sigma^{NLO}=\sigma^{0}+\sigma^{(2)}+\sigma^{(3)}.
\end{eqnarray}

\section{Numeric results and discussion}
In the following numerical evaluation, we present the results of
the cross section for the pseudoscalar Higgs boson production
associated with a single high-$p_T$ bottom quark via subprocess
$bg(\bar{b}g) \to A^0b(A^0\bar{b})$ at the LHC and Tevatron. At
the LHC, the $b$-jet is required to have a transverse momentum cut
$p_T(b)>30~GeV$ and a rapidity cut $|\eta(b)|<2.5$. At the
Tevatron, the $b$ tagging regions are taken to be $|\eta(b)|<2$
and $p_T(b)>15~GeV$. The SM parameters are taken as: $
m_t=174.3~GeV$, $m_Z = 91.188~GeV$, $m_{W}=80.419~GeV$ and
$\alpha_{EW} = 1/128$ \cite{pdg}. The factorization scale is taken
as $\mu_f=m_A/4$ and the renormalization scale is taken as
$\mu_r=m_A$. We use the one-loop formula for the running strong
coupling constant $\alpha_s$ with $\alpha_s(m_Z)=0.117$.

The relevant MSSM parameters in our calculation are: the
parameters $M_{\tilde{Q},\tilde{U},\tilde{D}}$ and $A_{t,b}$ in
squark mass matrices, the higgsino mass parameter $\mu$, the
masses of the gluino $m_{\tilde{g}}$ and the $A^0$ Higgs boson
$m_{A}$, the ratio of the vacuum expectation values of the two
Higgs doublets $\tan\beta$. The squark mass matrix is defined as
\begin{eqnarray}
{\cal M}_{\tilde{q}}^2 &=& \left( \begin{array}{cc} m^2_{\tilde{q}_L} & a_q m_q \\
a_q m_q & m^2_{\tilde{q}_R}
\end{array}\right)
\end{eqnarray}
with
\begin{eqnarray}
m^2_{\tilde{q}_L}&=&M_{\tilde{Q}}^2 + m_q^2 + m_Z^2 \cos 2\beta (I^q_3 - e_q \sin^2 \theta_W), \nb \\
m^2_{\tilde{q}_R}&=&M_{\{\tilde{U},\tilde{D}\}}^2 + m_q^2 + m_Z^2
\cos 2\beta e_q \sin^2 \theta_W   \nb \\
a_q&=&A_q-\mu \{\cot\beta,\tan\beta\},
\end{eqnarray}
for \{up, down\} type squarks. $I_3^q$ and $e_q$ are the third
component of the weak isospin and the electric charge of the quark
$q$. The chiral states $\tilde{q}_L$ and $\tilde{q}_R$ are
transformed into the mass eigenstates $\tilde{q}_{1}$ and
$\tilde{q}_{2}$:
\begin{equation}
\left( \begin{array}{cc} \tilde{q}_{1} \\ \tilde{q}_{2}
\end{array}
\right) = R^{\tilde{q}}\left( \begin{array}{cc} \tilde{q}_L \\
\tilde{q}_R \end{array} \right),~~R^{\tilde{q}} = \left(
\begin{array}{cc}\cos\theta_{\tilde{q}} &
 \sin\theta_{\tilde{q}} \\-\sin\theta_{\tilde{q}} & \cos\theta_{\tilde{q}} \end{array}
 \right).
\end{equation}
Then the mass eigenvalues $m_{\tilde{q}_{1}}$ and
$m_{\tilde{q}_{2}}$ are given by
\begin{eqnarray}
\left( \begin{array}{cc} m^2_{\tilde{q}_1} & 0 \\
0 & m^2_{\tilde{q}_2}\end{array}\right)=R^{\tilde{q}}{\cal
M}_{\tilde{q}}^2(R^{\tilde{q}})^{\dagger}
\end{eqnarray}
For simplicity, we assume $M_{\tilde{Q}}= M_{\tilde{U}}=
M_{\tilde{D}}=A_t=A_b=m_{\tilde{g}}$ collectively denoted by
$M_{SUSY}$.

In our calculation, we use the CTEQ5L parton distribution
functions\cite{lai}. The $\overline{\rm MS}$ bottom quark mass
$\overline{m}_b$ can be evaluated by using the next-leading order
formula \cite{msbar}. In the following equations, we use
$\overline{m}_b(Q)$ to denote the $\overline{\rm MS}$ bottom quark
mass.
\begin{eqnarray}
\overline{m}_b(Q) &=& U_5(Q,\overline{m}_b)\overline{m}_b(\overline{m}_b),~~~~~~{\rm for}~Q<m_t,  \nb \\
\overline{m}_b(Q) &=&
U_6(Q,m_t)U_5(m_t,\overline{m}_b)\overline{m}_b(\overline{m}_b),~~~~~~{\rm
for}~Q>m_t,
\end{eqnarray}
where $\overline{m}_b=\overline{m}_b(\overline{m}_b)=4.3~GeV$. The
evolution factor $U_f(f=5,6)$ is
\begin{eqnarray}
U_f(Q_2,Q_1)=\left(
\frac{\alpha_s(Q_2)}{\alpha_s(Q_1)}\right)^{d^{(f)}}
             [1+\frac{ \alpha_s(Q_1)-\alpha_s(Q_2)}{4 \pi} J^{(f)}], \nb \\
d^{(f)}=\frac{12}{33-2f},~~~~~~~~~~~~~~~J^{(f)}=-\frac{8982-504f+40f^2}{3(33-2f)^2}
\end{eqnarray}
In the supersymmetry limit, bottom quarks only couple to the
neutral Higgs doublet $H^0_1$. However, supersymmetry is broken
and the bottom quark will receive a small coupling to the Higgs
doublet $H_2^0$ from radiative corrections. Considering these
corrections, $A^0$ Higgs boson coupling to the bottom quarks is
given as \cite{vertex correct}
\begin{eqnarray}
\label{vertex2} -  \overline{m}_b \frac{g_w \tan\beta}{2
m_W}\gamma_5~\to~- \frac{\overline{m}_b}{1+\Delta_b}\frac{g_w
\tan\beta}{2 m_W}\gamma_5
\end{eqnarray}
The explicit form of $\Delta_b$ at one-loop is given by
\cite{dltb1,dltb2,dltb3}
\begin{eqnarray}
\Delta_b = \frac{2 \alpha_s}{3 \pi} m_{\tilde{g}}\tan \beta
I(m_{\tilde{b}_1},m_{\tilde{b}_2},m_{\tilde{g}})+\frac{h_t^2}{16
\pi^2}A_t\mu\tan\beta I(m_{\tilde{t}_1},m_{\tilde{t}_2},\mu),
\end{eqnarray}
where $h_t=\frac{\sqrt{2} m_t}{v \sin \beta},~v=246GeV$ and the
function $I$ is given by
\begin{eqnarray}
I(a,b,c)=\frac{1}{(a^2-b^2)(b^2-c^2)(a^2-c^2)}(a^2b^2\log
\frac{a^2}{b^2} + b^2c^2\log \frac{b^2}{c^2} + c^2a^2\log
\frac{c^2}{a^2}).
\end{eqnarray}

Fig.5 shows that our NLO-QCD result does not depend on the
arbitrary cutoffs $\delta_s$ and $\delta_c$ by using the two
cutoff phase space slicing method. The two-body($\sigma^{(2)}$)
and three-body($\sigma^{(3)}$) cross sections and the NLO cross
section ($\sigma^{NLO}$) at the LHC, are shown as the functions of
the soft cutoff $\delta_s$ with the collinear cutoff
$\delta_c=\delta_s/50$ . The supersymmetric parameters are taken
as $\mu=m_A=200~GeV$, $M_{SUSY}=500~GeV$ and $\tan\beta=4$. We can
see the NLO cross section $\sigma^{NLO}$ is independent of the
cutoffs. In the following numerical calculations, we take
$\delta_s=10^{-4}$ and $\delta_c=\delta_s/50$.

\FIGURE{
  \leavevmode
  \begin{tabular}{cc}
    \epsfxsize=25em
      \epsffile[100 60 500 420]{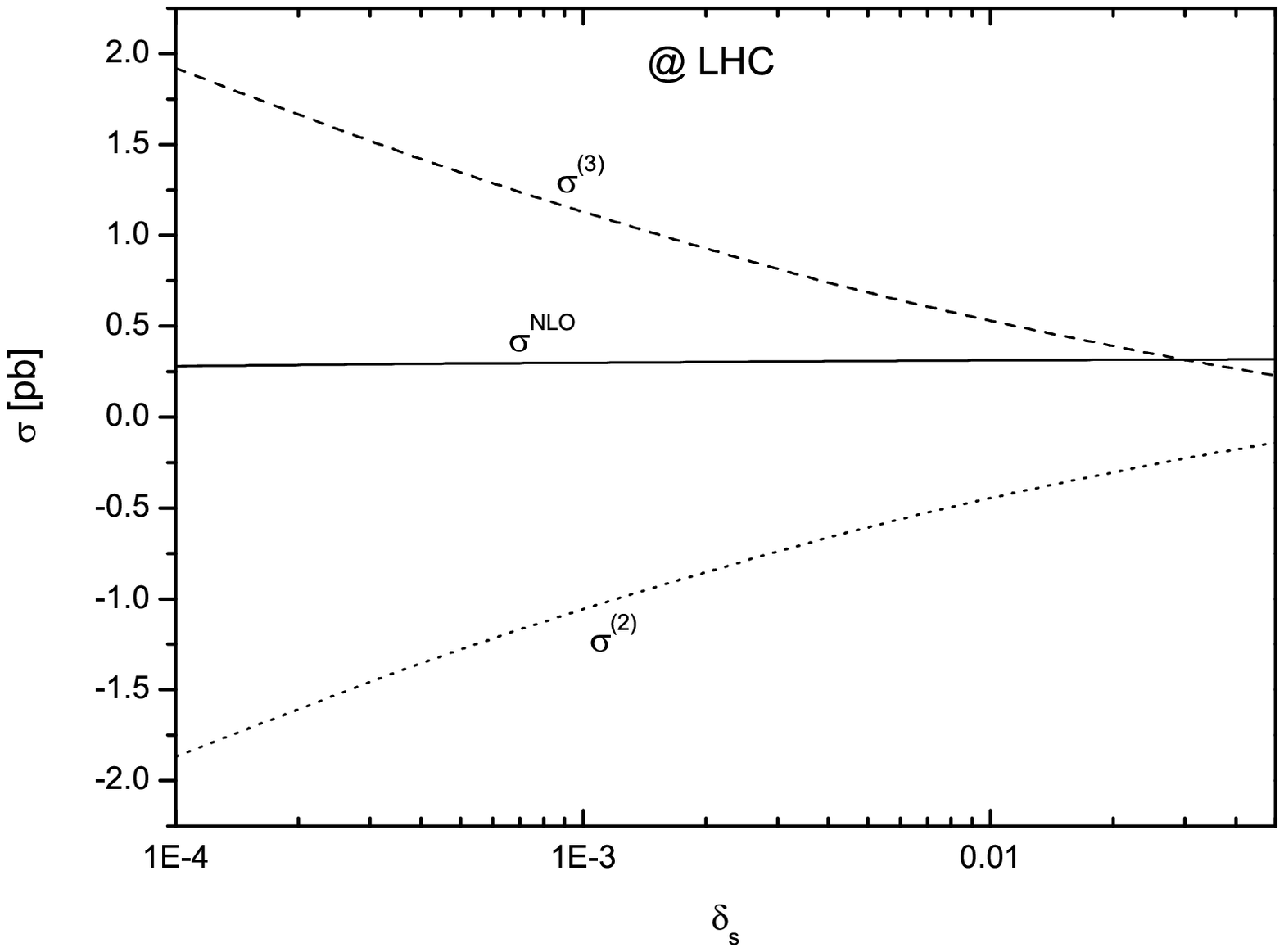}
  \end{tabular}
  \caption[]{\label{fig::fig5}\sloppy
    Dependence of the cross sections for the $A^0b$
production at the LHC on the cutoff $\delta_s$ with
$\delta_c=\delta_s/50$. }
  }

Fig.6 shows the dependence of the LO and NLO cross sections of the
process $pp$ (or $p\bar{p}) \to bg(\bar{b}g) \to
A^0b(A^0\bar{b})+X$ at the LHC and Tevatron on the mass of $A^0$
Higgs boson($m_A$). Here we take $\mu=200~GeV$, $M_{SUSY}=500~GeV$
and $\tan\beta=4,~15,~30$. The relative NLO-QCD corrections are
about $40\%\sim 50\%$ at the LHC and about $45\% \sim 80\%$ at the
Tevatron when $m_A$ varies from $200~GeV$ to $800~GeV$ for all the
values of $\tan\beta$ we have taken.

\FIGURE{
  \leavevmode
  \begin{tabular}{cc}
    \epsfxsize=25em
      \epsffile[120 60 500 350]{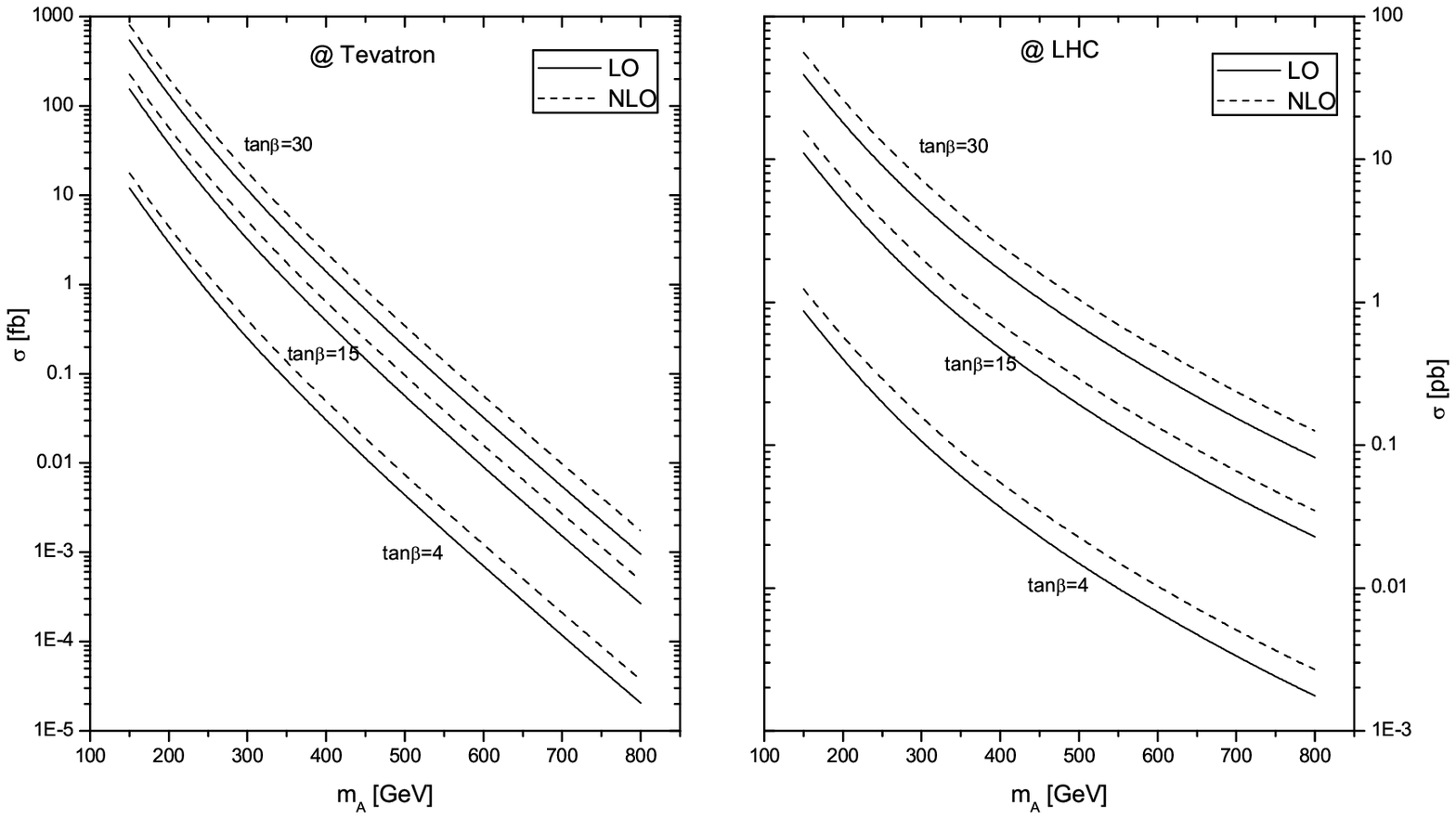}
  \end{tabular}
  \caption[]{\label{fig::fig6}\sloppy
    The dependence of the cross sections of process $pp$
(or $p\bar{p}) \to bg(\bar{b}g) \to A^0b(A^0\bar{b})+X$ on the
$m_A$ with $\tan\beta=4,15,30$ at the LHC and Tevatron.}
  }

Fig.7 shows the cross sections of the process $pp$ (or $p\bar{p})
\to bg(\bar{b}g) \to A^0b(A^0\bar{b})+X$ at the LHC and Tevatron
as the functions of the ratio of the expectation vacuum values
$\tan{\beta}$. We take $\mu=200~GeV$, $M_{SUSY}=500~GeV$ and
$m_A=200,~500,~800~GeV$. Since the coupling between $A^0$ Higgs
boson and bottom quarks is greatly enhanced with large
$\tan\beta$(see Eq.(\ref{vertex1}), Eq.(\ref{vertex2})), the cross
section of process $pp$ (or $p\bar{p}) \to bg(\bar{b}g) \to
A^0b(A^0\bar{b})+X$ at the LHC and Tevatron can be rather large.
We can see that the cross section of the process $pp$ (or
$p\bar{p}) \to bg(\bar{b}g) \to A^0b(A^0\bar{b})+X$ can reach
dozens of pico bar(or about $1~pb$) when $m_A=200~GeV$ nad
$\tan\beta=40$(or when $m_A=500~GeV$ and $\tan\beta=40$) at the
LHC, while the cross section can be $200~fb$ when $m_A=200~GeV$
and $\tan\beta=40$ at the Tevatron.

\FIGURE{
  \leavevmode
  \begin{tabular}{cc}
    \epsfxsize=25em
      \epsffile[120 60 500 350]{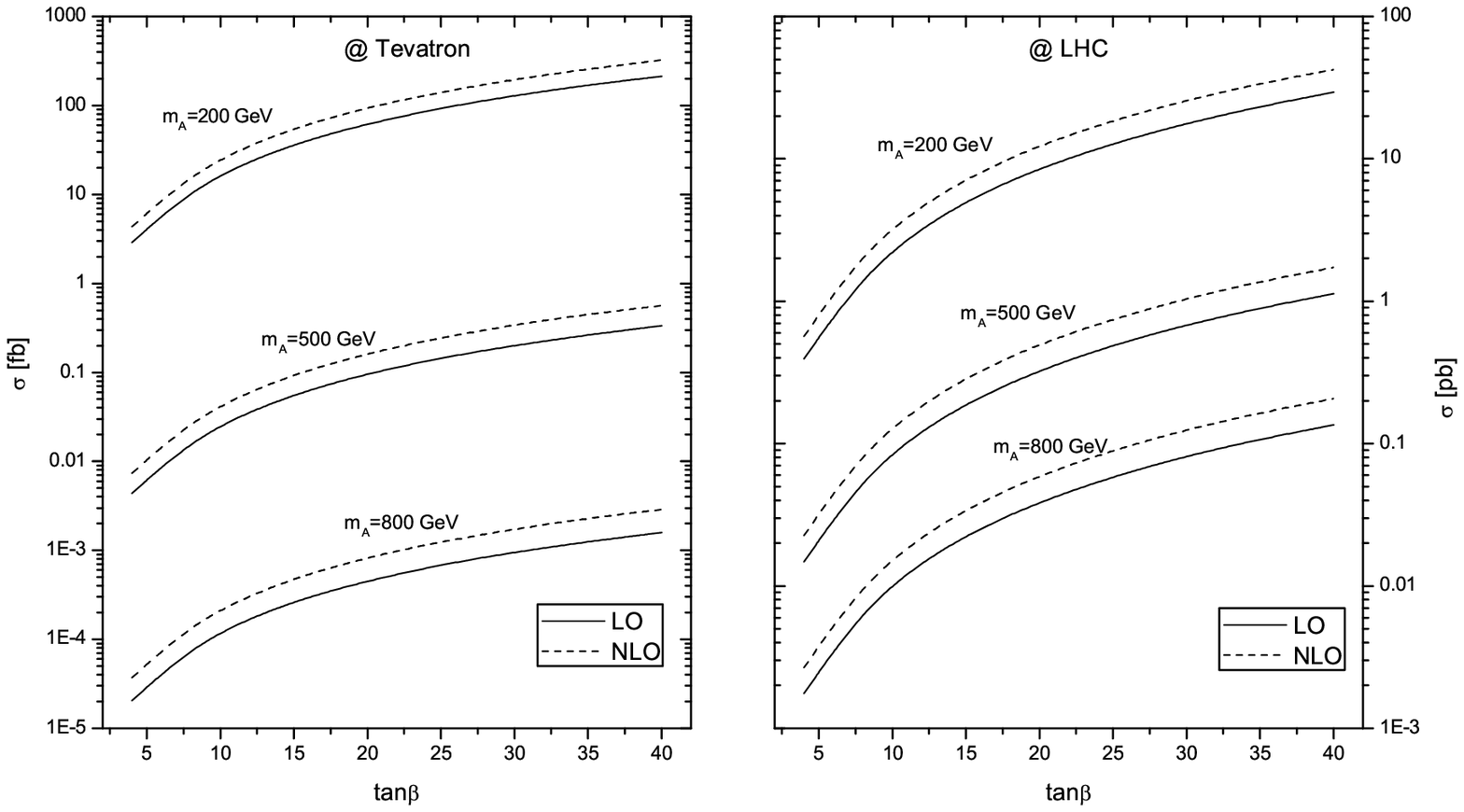}
  \end{tabular}
  \caption[]{\label{fig::fig7}\sloppy
    The dependence of the cross sections of process $pp$
(or $p\bar{p}) \to bg(\bar{b}g) \to A^0b(A^0\bar{b})+X$ on
$\tan\beta$ with $m_A=200,500,800GeV$ at the LHC and Tevatron.}
  }

In order to study the decoupling behavior of SUSY QCD correction,
we push $M_{SUSY}$ to a large value. The relative SUSY QCD
correction is defined as
\begin{eqnarray}
\Delta_{SQCD}=\frac{\delta \sigma^{SQCD}}{\sigma^0},
\end{eqnarray}
where $\delta \sigma^{SQCD}$ is the cross section correction
contributed by the SUSY QCD diagrams shown in Fig.3. In Fig.8, we
depict the relative SUSY QCD correction $\Delta_{SQCD}$ as the
functions of $M_{SUSY}$ at the LHC. The relative SUSY QCD
correction is small (about $2.2\% \sim 2.5\%$) but not vanishing
with $M_{SUSY}$ up to $2~TeV$.

\FIGURE{
  \leavevmode
  \begin{tabular}{cc}
    \epsfxsize=25em
      \epsffile[120 60 500 430]{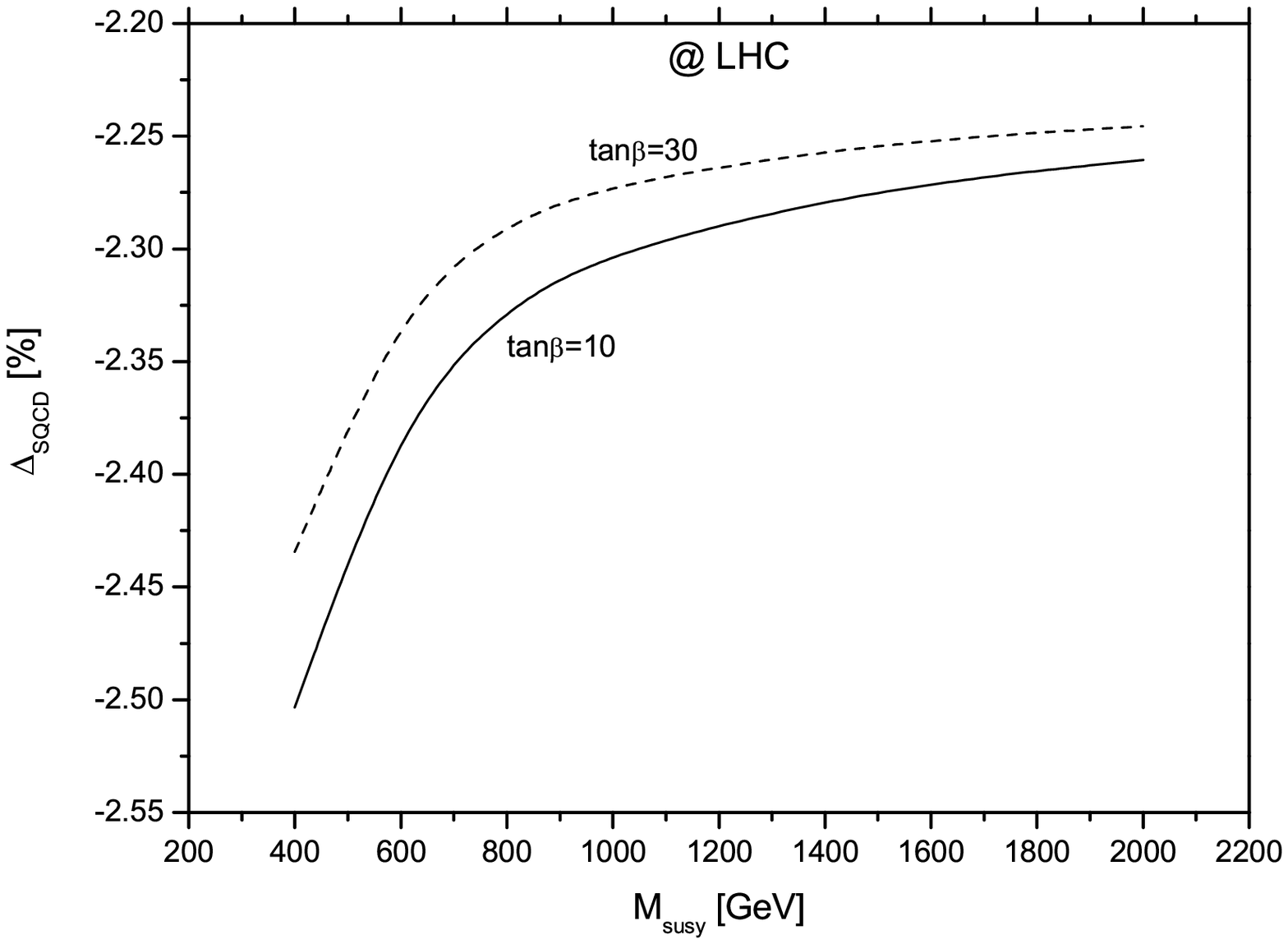}
  \end{tabular}
  \caption[]{\label{fig::fig8}\sloppy
    The dependence of $\Delta_{SQCD}$  of process $pp \to
bg(\bar{b}g) \to A^0b(A^0\bar{b})+X$ on $M_{SUSY}$ at the LHC.}
  }

To analyze the scale dependence of the cross sections, We
introduce the ratio of the cross section at scale $\mu$ and the
cross section at scale $\mu=m_{A}$ and depict the $\sigma(\mu)/
\sigma(\mu=m_{A})$ as a function of $\mu/m_{A}$ at the LHC in
Fig.9. For the solid lines, we fix the renormalization scale
$\mu_r=m_A$ and only show factorization scale $\mu_f$ dependence
of the cross sections. For dashed lines, $\mu_f$ and $\mu_r$ are
taken to be identical, vary from $m_A/5$ to $2m_A$. The scale
variation of the NLO-QCD cross section may be serves as an
estimate of the remaining theoretical uncertainty of the high
order corrections. Fig.9 shows that it is evident that the
one-loop NLO-QCD corrections reduce the LO scale dependence for
both the cases.

\FIGURE{
  \leavevmode
  \begin{tabular}{cc}
    \epsfxsize=25em
      \epsffile[120 60 500 430]{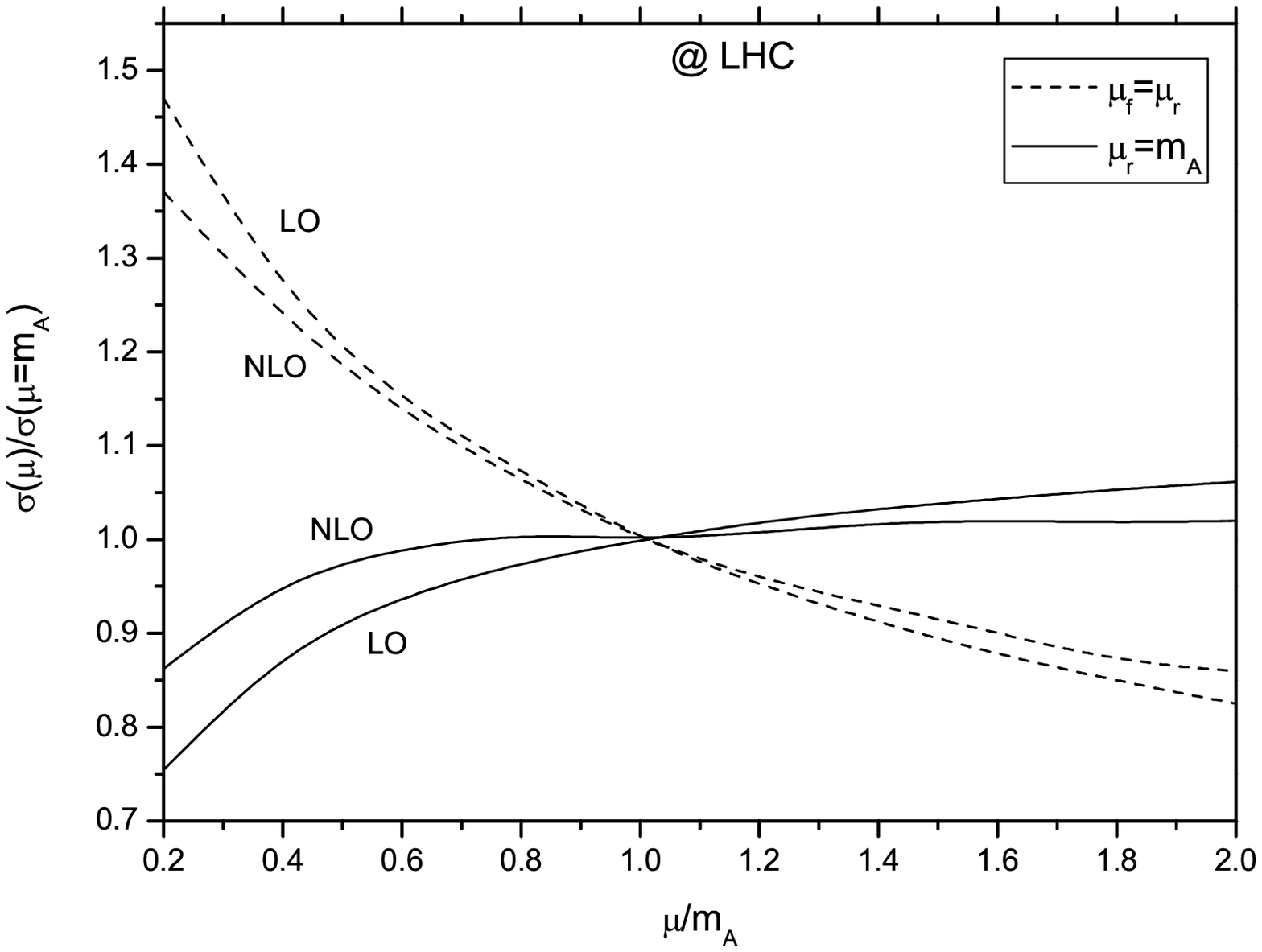}
  \end{tabular}
  \caption[]{\label{fig::fig9}\sloppy
    The variation of the $\sigma(\mu) / \sigma(\mu=m_{A})$ with the
ratio $\mu/m_{A}$ of process $pp \to bg(\bar{b}g) \to
A^0b(A^0\bar{b})+X$ at the LHC.}
  }

In summary, we have computed the production of pseudoscalar Higgs
boson $A^0$ associated with a single high-$p_T$ bottom quark via
subprocess $bg(\bar{b}g) \to A^0b (A^0\bar{b})$ including the
NLO-QCD corrections in the MSSM at the LHC and Tevatron. We find
that due to the enhancement of the Yukawa coupling strength of the
bottom quarks with $A^0$ Higgs bosons at large $\tan{\beta}$, the
cross section of the $pp$ (or $p\bar{p}) \to A^0b(A^0\bar{b})$ can
reach dozens of pico bar at the LHC and hundreds of fermi bar at
the Tevatron. The NLO-QCD corrections vary between $40\% \sim
50\%$ at the LHC and $45\% \sim 80\%$ at the Tevatron
respectively, in the parameter space we have chosen.

\paragraph{Acknowledgments.}
This work was supported in part by the National Natural Science
Foundation of China and a grant from the University of Science and
Technology of China.

\end{document}